\documentclass[%
 aip,
 amsmath,amssymb,
 reprint,%
]{revtex4-1}
\usepackage{multirow}
\usepackage{graphicx}
\usepackage{dcolumn}
\usepackage{bm}
\usepackage{color}
\usepackage{booktabs} 
\usepackage[utf8]{inputenc}
\usepackage[T1]{fontenc}
\usepackage{mathptmx}
\usepackage{etoolbox}

\makeatletter
\def\@email#1#2{%
 \endgroup
 \patchcmd{\titleblock@produce}
  {\frontmatter@RRAPformat}
  {\frontmatter@RRAPformat{\produce@RRAP{*#1\href{mailto:#2}{#2}}}\frontmatter@RRAPformat}
  {}{}
}%
\makeatother
\begin{document}

\preprint{AIP/123-QED}

\title{Unified signal response for stochastic resonance in bistable systems }
\author{Cong Liu}
\email[Author to whom correspondence should be addressed:]{cliu@hebtu.edu.cn}
\affiliation{ College of Physics and Hebei Key Laboratory of Photophysics Research and Application, Hebei Normal University, Shijiazhuang, Hebei 050024, China }
\affiliation{ 
Lanzhou Center for Theoretical Physics, Key Laboratory of Theoretical Physics of Gansu Province, and Key Laboratory of Quantum Theory and Applications of MoE, Lanzhou University, Lanzhou, Gansu 730000, China
}%

\affiliation{ 
Institute of Computational Physics and Complex Systems,
Lanzhou University, Lanzhou, Gansu 730000, China
}%
\author{Xin-Ze Song}%
\email{aaasongxinze@163.com}
\affiliation{ 
Lanzhou Center for Theoretical Physics, Key Laboratory of Theoretical Physics of Gansu Province, and Key Laboratory of Quantum Theory and Applications of MoE, Lanzhou University, Lanzhou, Gansu 730000, China
}%
\affiliation{ 
Institute of Computational Physics and Complex Systems,
Lanzhou University, Lanzhou, Gansu 730000, China
}%
\author{Zhi-Xi Wu}
\email{wuzhx@lzu.edu.cn}

\affiliation{ 
Lanzhou Center for Theoretical Physics, Key Laboratory of Theoretical Physics of Gansu Province, and Key Laboratory of Quantum Theory and Applications of MoE, Lanzhou University, Lanzhou, Gansu 730000, China
}%
\affiliation{ 
Institute of Computational Physics and Complex Systems,
Lanzhou University, Lanzhou, Gansu 730000, China
}%
\author{Guo-Yong Yuan}
\affiliation{ College of Physics and Hebei Key Laboratory of Photophysics Research and Application, Hebei Normal University, Shijiazhuang, Hebei 050024, China }
\date{\today}

\begin{abstract}
The phenomenon of stochastic resonance, wherein the stimulus-response of a system can be maximized by an intermediate level of noise, has been extensively investigated through linear response theory. 
As yet a unified response-noise or response-frequency formula embracing diverse factors, such as noise color, damping coefficients, and coupling, is still lacking. 
In the present work, we theoretically investigate the benefit roles of Gaussian white noise and Ornstein-Uhlenbeck noise on the signal amplification of systems ranging from a single overdamped bistable particle to the mean-field coupled underdamped Duffing oscillators and severally deduce their signal response expressions. 
We find that the formulas of signal response in these different cases can be reduced to a uniform Lorentz function form. 
Furthermore, based on the general expression, we explain explicitly the role of driving frequency, coupling and the noise color on stochastic resonance. 
Our results contribute to a deep theoretical understanding of stochastic resonance in bistable systems.
\end{abstract}

\maketitle

\begin{quotation}
Noise or fluctuations are ubiquitous in natural and man-made systems. The pioneering study on stochastic resonance significantly encourages people to make use of noise to amplify signals. Despite the theoretical investigation on such a resonance phenomenon has never stopped, to our best knowledge, a general signal response expression applied to diverse circumstances ranging from a single overdamped bistable particle driven by the Gaussian white noise to the mean-field coupled underdamped Duffing elements perturbed by the Ornstein-Uhlenbeck noise is rarely reported. In the present work, based on linear response theory, we deduced respectively the signal response expressions for these distinct cases, and found that the individual signal responses can be concluded into a uniform Lorentz function class. These results integrate the independent theoretical results and thus contribute to a deep understanding of stochastic resonance in bistable systems.
\end{quotation}

\section{\label{sec:level1}INTRODUCTION}
Noise is widespread in natural and artificial systems~\cite{Benzi1981,Nicolis1981,Lucarini2019,Gammaitoni1998,HPZhang2022,Burada2008,Lindner2004,McDonnell2011,Zhu2021,Marius2024,Zhang2019}, i.e., annual fluctuations in solar radiation~\cite{Benzi1981,Nicolis1981,Lucarini2019}, thermodynamic fluctuations in molecular motion~\cite{Gammaitoni1998,HPZhang2022,Burada2008}, electrical or chemical noise in neural activity and cardio-motility
~\cite{Gammaitoni1998,Sag2007,Lindner2004,McDonnell2011,Zhu2021,Marius2024}, disturbance of the environment in the power grid~\cite{Zhang2019,Zhang2020,Anvari2020}, among others. Like the annoyed hiss disturbing the communication quality of radio, noise is conventionally viewed as the destroyer in signal processing and transmission~\cite{Gammaitoni1998}.
The question concerning why such a purely "negative" element is so  
widespread has been one of the core topics in nonlinear dynamics and statistical physics in the past decades.
The notion of stochastic resonance (SR)~\cite{Benzi1981,Nicolis1981}, in which noise can play a positive role in inducing stimulus-response dynamics, breaks our conventional wisdom and has been verified in almost all scientific disciplines~\cite{Gammaitoni1998,McNamara1988,Anishchenk1999,Lucarini2019,Karabalin2011,Badzey2005,Zaks2005,Li2018,Greenwood2000,Jung1991,Wang2009}, see the acclaimed review~\cite{Gammaitoni1998} and references therein.
The dynamical archetype of SR can be depicted by a heavy Brownian particle moving in a symmetrical bistable potential. Noise occasionally activates the interwell hopping of the particle and the hopping rate can be quantitatively characterized by the Kramers escape rate~\cite{Gammaitoni1998,Lucarini2019}. Without noise, the subthreshold driving signal fails to activate the particle to jump into the other potential. For a strong noise, particle motion submerges in the noise environment. While for an intermediate degree of noise (especially the noise-induced hopping matches the driving force, that is, the hopping rate is twice the driving frequency) the averaged oscillation with a maximum amplitude can be observed.

On the one hand, since the overwhelming majority of interest in SR is concentrated on weak external stimulus input, the powerful tool of linear response theory (LRT) is frequently used to obtain the theoretical stimulus response~\cite{Gammaitoni1998,Hanggi1982}. In LRT, the magnitude of the system response is proportional to the driving amplitude, say $\langle x(t)\rangle_{t}=\chi(D,\omega) A\sin(\omega t)$, in which the symbol $\chi(D,\omega)$ stands for the response function or susceptibility. Based on the fluctuation-dissipation theorem, the susceptibility can be further obtained by the self-correlation of the system when the external driving is switched off~\cite{Gammaitoni1998,Hanggi1982}. So far, enormous effort has been devoted to utilizing such a theory to investigate SR. As striking examples, Dykman and coworkers gave the qualitative and quantitative understanding of SR in the system of a single bistable oscillator, and further corroborated the phase shift behavior in SR~\cite{Dykman1990,Dykman1992}. At the same time, McNamara \emph{et al}. proposed a key yet powerful simplification of SR in the two-state theory~\cite{McNamara1988,McNamara1989}. Jung \emph{et al}. proposed the theoretical description of SR without restrictions on driving amplitude and frequency~\cite{Jung1991,Jung1993}. Hu \emph{et al}. verified for the first time that the sensitive frequency dependence can also emerge in SR in globally coupled continue cells with excitatory and inhibitory properties~\cite{Hu1996,Zhang1998}. Morillo \emph{et al}. investigated the signal response of mean-filed coupled bistable elements with a numerical response function~\cite{Morillo1995}. Frank deduced the response function of the Desai-Zwanzig model~\cite{Frank2004}. Pascual and collaborators examined the  validity of LRT and found that the output signal amplitude significantly deviates from the LRT prediction when the driving frequency is sufficiently low~\cite{Casado-Pascual2002}. Kang \emph{et al}. explored SR and bifurcation of the globally coupled underdamped Duffing oscillators~\cite{Liu2019}. Lucarini proposed a general quasipotential for describing SR in noisy $N$-dimensional nonequilibrium systems possessing two metastable states~\cite{Lucarini2019}. Besides, the theoretical investigation interest has also been extended to the multiplicative noise situations~\cite{Zaikin2003,Qiao2016}. Qiao \emph{et al}. systematically investigated the influence of the potential asymmetries on SR under the action of both multiplicative and additive noise~\cite{Qiao2016}.

On the other hand, as colored noise is more common than white noise in the physical world~\cite{Hanggi1994}, LRT is utilized to explore SR induced by colored noise as well. For instance,  H\"{a}nggi and coworkers demonstrated that the noise color, namely the correlation time, weakens the magnitude of SR for a single overdamped bistable oscillator~\cite{Hanggi1993}. Kang and participators proposed a semi-analytic method for quantitatively investigating the long-time ensemble dynamics of the mean-field coupled overdamped bistable oscillators driven by Ornstein-Uhlenbeck (OU) noise~\cite{Kang2008}.  Despite these findings, the theoretical explorations on SR are relatively fragmented. Whether there is a general response-noise formula embracing diverse factors, like noise color, damping coefficients and coupling, is still unknown.

In the present work, based on LRT, we comprehensively investigate the signal response of systems ranging from a single overdamped bistable oscillator to the mean-field coupled underdamped Duffing elements driven by Gaussian white noise and OU noise, respectively. We derive the signal response expressions in these distinct situations and find that these signal response results can be reduced to a general theoretical formula with the Lorentz function form. 
Moreover, we explicitly explain the role of the external driving frequency, the mean-field coupling and the noise color on SR through the general signal response expression.
These results integrate the independent theoretical results and thus contribute to a deep understanding of stochastic resonance in bistable systems.

The remainder of this paper is organized as follows. In Sec.\ref{Sec2A}, we deduce respectively the theoretical signal response expressions of a single overdamped bistable particle and the mean-field coupled overdamped bistable systems with respect to Gaussian white noise and OU noise. Similar investigations on the underdamped Duffing counterparts are shown in Sec.\ref{Sec2B}. Based on the general signal response, the influence of diverse factors on SR is clearly provided in Sec.\ref{Sec2C}. Finally, a brief summary and discussions of our main results are given in Sec.\ref{Sec2D}.

\section{SR For overdamped bistable model}\label{Sec2A}

\subsection{Single oscillator under Gaussian white noise }
Since the main objective of the present work is to deduce a general signal response formula embracing diverse factors of bistable systems through a unified LRT viewpoint, the case of a single overdamped particle driven by the Gaussian white noise and the weak sinusoidal signal is both the archetype of SR and the starting point of the present work, we first pay attention to the signal response in this case. On the other hand, as the frequency of the external signal is the core factor in a resonance-like phenomenon, we also focus on the influence of the driving frequency on signal response. The dynamics of a single oscillator driven by the Gaussian white noise and weak signal can be represented by the Langevin equation 
\begin{equation}\label{eq:singlebistable}
\dot{x}=x-x^{3}+A\sin(\omega t)+\xi(t), 
\end{equation}
in which $x$ is the state variable. The symbol $\xi(t)$ stands for the Gaussian white noise, of which the first and second moments are $\langle \xi(t) \rangle=0$ and $\langle \xi(t)\xi(t')\rangle =2D\delta(t-t')$, respectively. Therefore, $D$ measures the intensity of noise. Additionally, the power spectrum of $\xi(t)$, $S_{\textrm{White}}=\int_{-\infty}^{\infty}\langle\xi(t)\xi(t')\rangle e^{-i\omega t}dt=2D$, pervades the whole frequency band. As a result, one needs an infinite amount of energy to generate white noise. Hereinafter, we utilize the the spectral amplification factor~\cite{Jung1991,Jung1993},
\begin{equation}\label{eq:Amplification factor}
\eta_{\textrm{single}}=\frac{4}{A^{2}} \left| \langle e^{i\omega t}x(t)\rangle \right|^{2}, 
\end{equation}
to measure the signal response of the single bistable particle.

The Fokker-Planck equation of Eq.~(\ref{eq:singlebistable}) can be recast as  
\begin{eqnarray}\label{eq:2}
\frac{\partial W(x,t)}{\partial t}&=&-\frac{\partial}{\partial x}\left\{(x-x^{3})W(x,t)\right\}+D\frac{\partial^{2}}{\partial x^{2}}W(x,t)\nonumber\\
&-&A\sin(\omega t)\frac{\partial}{\partial x}W(x,t)\nonumber\\
&=&L_{0}W(x,t)+L_{1}W(x,t),
\end{eqnarray}
where $W(x,t)$ depicts the state probability density function at time $t$ of Eq.~(\ref{eq:singlebistable}), besides, $L_{0}=-\partial (x-x^{3})/\partial x+\partial^{2}D/\partial x^{2}$, and $L_{1}=-\partial A\sin(\omega t)/\partial x=-\partial F_{\textrm{ext}}(t)/\partial x $
stand for the Fokker-Planck operator of the vanished stimulus case and  the stimulated one, respectively. Equation ~(\ref{eq:2}) specifies that adding a periodical stimulus on the system equals to adding a periodical perturbation that correlates to the state variable on the original Fokker-Planck equation.

The steady-state solution of the Fokker Planck equation in the absence of the external stimuli can be obtained as
\begin{equation}\label{eq:5}
W_{\textrm{st}}(x)=\frac{Z}{D}\exp\left[\frac{1}{D}(\frac{1}{2}x^{2}-\frac{1}{4}x^{4})\right],
\end{equation}
where $Z$ is the normalization factor of Eq.~(\ref{eq:5}).
Additionally, once the external driving adds, a new term in the probability density function (PDF), $W_{\textrm{ext}}(x,t)$ emerges, and as a result, the PDF of the whole system can be further read as 
\begin{equation}\label{eq:6}
W(x,t)=W_{\textrm{st}}(x)+W_{\textrm{ext}}(x,t),
\end{equation}
and the evaluation of expression for $W_{\textrm{ext}}(x,t)$ is 
\begin{equation}\label{eq:7}
W_{\textrm{ext}}(x,t)=\int_{-\infty}^{t}e^{L_{0}(t-t')}L_{1}(t')W_{\textrm{st}}(x)dt'.
\end{equation}
Based on the LRT, the averaged state variable can be fragmented into two parts:
\begin{equation}\label{eq:8}
\langle x(t)\rangle=\langle x_{0}\rangle+\langle x_{1}(t)\rangle,
\end{equation}
in which $\langle x_{0} \rangle$ and $\langle x_{1} \rangle$ are the time-averaged variables without and within the influence of the external stimulus, respectively. The Eq.~(\ref{eq:8}) can be further recast as 
\begin{equation}\label{eq:9}
\langle x_{1}(t)\rangle=\int xW_{\textrm{ext}}(x,t)dx.
\end{equation}
Inserting Eq.~(\ref{eq:7}) into Eq.~(\ref{eq:9}), we obtain the averaged variable under the external stimulus 
\begin{eqnarray}\label{eq:10}
\left\langle x_{1}(t)\right\rangle&=&\int xW_{\textrm{ext}}(x,t)dx\nonumber\\
&=&-\int_{-\infty}^{t}F_{\textrm{ext}}(t')\left[\int_{x}xe^{L_{0}(t-t')}\frac{\partial}{\partial x}W_{\textrm{st}}(x)dx\right]dt'\nonumber\\
&=&\int_{-\infty}^{t}\chi(t-t')F_{\textrm{ext}}(t')dt',
\end{eqnarray}
in which the symbol  $\chi(t-t')$ represents the response function, also known as susceptibility, demonstrating the intensity coefficient for system responses to the weak stimulus at the moment $t$. Additionally, the response function is capable of the properties
\begin{equation}\label{eq:divide}
\chi(t-t')=\begin{cases}
\begin{array}{c}
\textrm{$-\int_{x}xe^{L_{0}(t-t')}\frac{\partial}{\partial x}W_{\textrm{st}}(x)dx,$}\\
\textrm{$0$,}
\end{array} & \begin{array}{c}
\textrm{ $t\ge t'$},\\
\textrm{$t< t'$}.
\end{array}\end{cases}
\end{equation}
As we mentioned in the preceding section, the product $\chi(t)F_{\textrm{ext}}(t)$ suggests that the relationship between system response and external stimulus is linear approximation. 

The transition probability of Eq.~(\ref{eq:singlebistable}) from state $(x,t)$ to state $(x',t')$ can be represented by
\begin{equation}\label{eq:13}
P(x,t|x',t')=e^{L_{0}(t-t')}\delta(x-x'),
\end{equation}
and further
\begin{equation}\label{eq:14}
P(x,t|x',t')x'=e^{L_{0}(t-t')}x.
\end{equation}
Bring Eq.~(\ref{eq:13}) into Eq.~(\ref{eq:divide}), and neglecting the higher-order terms, we have the response function
\begin{eqnarray}\label{eq:15}
\chi(t-t')&=&-\int_{x}xe^{L_{0}(t-t')}\frac{\partial}{\partial x}W_{\textrm{st}}(x)dx\nonumber\\
&=&-\frac{1}{D}\int_{x}\int_{x'}xx'P(x,t|x',t')dxdx'\nonumber\\
&=&-\frac{1}{D}\frac{d}{dt'}\langle x(t)x(t')\rangle_{0},
\end{eqnarray}
where $\langle...\rangle_{0}$ denotes the stationary average of the unperturbed process~\cite{Gammaitoni1998,Hanggi1982}.
Actually, the correlation function $\langle x(t)x(t')\rangle_{0}$ cannot be analytically calculated, we thus pay attention to an approximation analysis~\cite{Jung1991}. 
Based on the Kramers escape theory, the probability of the overdamped Brownian particle escapes from the left well to the right one can be written as~\cite{Gammaitoni1998} 
\begin{equation}\label{eq:16}
\lambda=\frac{1}{2\pi}\sqrt{V''(x_{s})\mid V''(x_{u})\mid } e^{-\frac{\Delta V}{D}},  
\end{equation}
in which $x_{s}=\pm 1$, $x_{u}=0$, $V(x)$ and $\Delta V$ are the potential energy function and the height of the potential barrier separating the two minima, respectively. The value of Kramers escape rate of Eq.~(\ref{eq:singlebistable}) is thus
\begin{equation}\label{eq:17}
\lambda=\frac{1}{\sqrt{2}\pi}e^{-\frac{1}{4D}}.  
\end{equation}
Moreover, the time correlation function can be viewed as the location correlation of the Brownian particle between the two moments $t$ and $t'$ in the potential well. The correlation function can be further rewritten as
\begin{equation}\label{eq:correlationfunction}
\langle x(t)x(t')\rangle_{0}=\int_{x}\int_{x'}xx'P(x,t|x',t')P(x',t')dxdx',
\end{equation} in which $P(x',t')$ is the probability of the particle in state $x'$ at the moment $t'$, and $P(x,t|x',t')$ stands for the transition probability from $x'$ at the moment $t'$ to $x$ at time $t$.  Without loss of generality, we take into account that the double-well potential has only two states which correspond to the two stable states $x_{s}$ and $x_{u}$, respectively. If the particle lies in the position $x_{s}$ at the moment $t$, then the probability that the particle arrives at the other one $x_{u}$ at the moment $t'$ is $\lambda$. As a consequence, the two probabilities above can be approximatively recast as
$P(x',t')=2^{-1}[\delta(x'-x_{u})+\delta(x'+x_{u})]$ and $P(x,t|x',t')=2^{-1}[\delta(x-x_{u})+\delta(x+x_{u})][1\pm e^{-\lambda t}]$, and thereby the correlation function reads as 
\begin{equation}\label{eq:19}
\langle x(t)x(t')\rangle_{0}=\langle x(t)^{2} \rangle_{0} e^{-\lambda t'}.
\end{equation}
Such a correlation function can be also obtained by the two-mode approximation~\cite{Jung1993}.
Inserting the correlation function into response function, we have the response function 
\begin{equation}\label{eq:20}
\chi(t-t')=\lambda\frac{\langle x(t)^{2}\rangle_{0}}{D}e^{-\lambda t'}.   
\end{equation}
Since the system response is 
\begin{equation}\label{eq:21}
\langle x_{1}(t)\rangle=\int_{-\infty}^{t}\chi(t-t')F_{\textrm{ext}}(t') dt',
\end{equation}
of which the Fourier transform can be written as 
\begin{equation}\label{eq:22}
\langle\widetilde{x}_{1}(\omega)\rangle=\widetilde{\chi}(\omega)\widetilde{F}_{\textrm{ext}}(\omega),
\end{equation}
where 
\begin{equation}
\widetilde{\chi}(\omega)=\int_{0}^{\infty}\chi(t-t')e^{i\omega t'} dt'
=\frac{\langle x(t)^{2} \rangle_{0}}{D}\left(\frac{-\lambda}{i\omega-\lambda}\right).    
\end{equation}\label{eq:23}
Then the signal response can be recast as  
\begin{equation}\label{eq:LRTsinglebistable}
\eta_{\textrm{single}}(\omega)=\mid\widetilde{\chi}(\omega)\mid^{2}=\left(\lambda\frac{\langle x(t)^{2} \rangle_{0}}{D}\right)^{2}\left(\frac{1}{\omega^{2}+\lambda^{2}}\right).    
\end{equation}
\begin{figure}
\centering
\includegraphics[width=1.0\columnwidth]{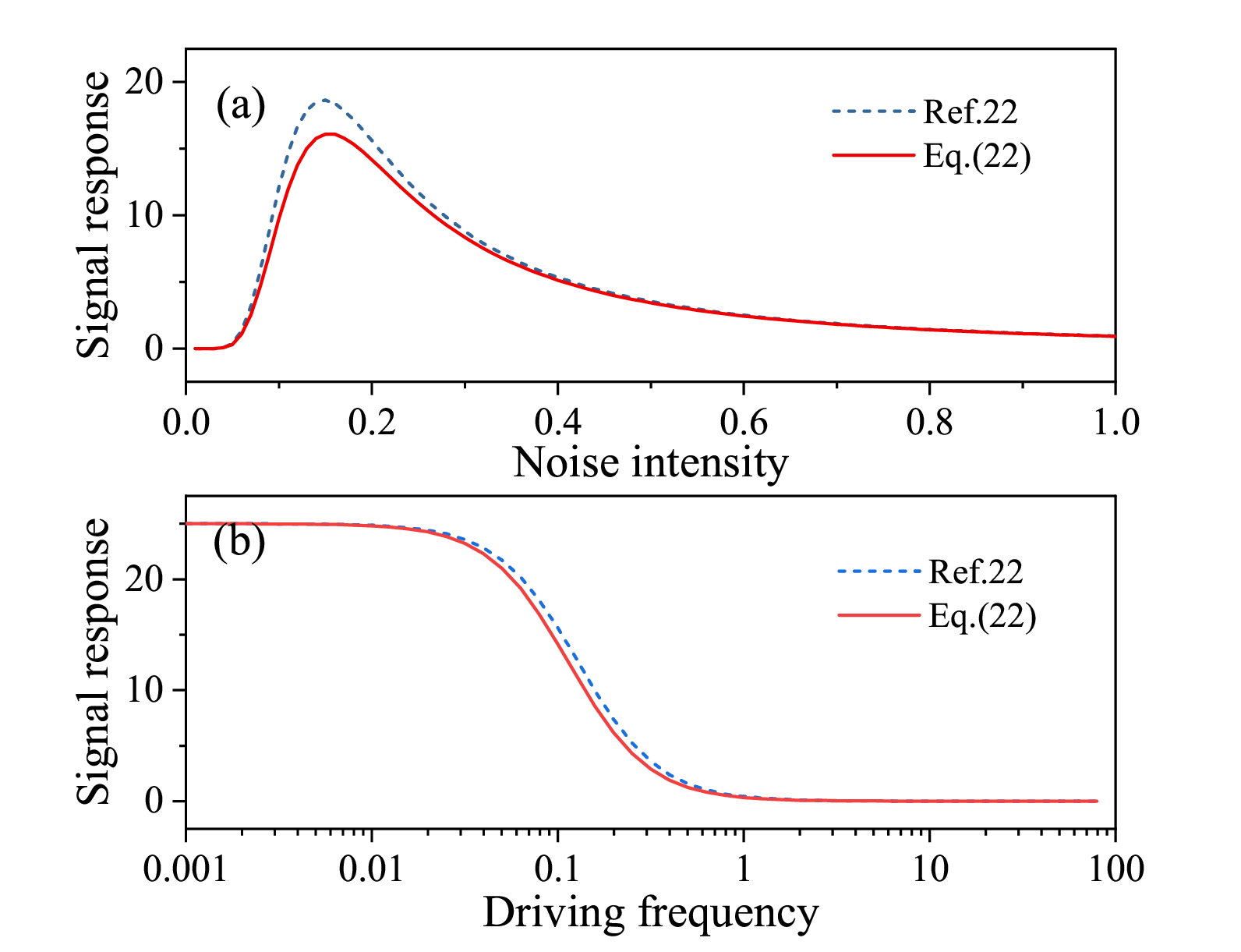}
\caption{(a) Signal response versus the noise intensity in system of a single overdamped bistable oscillator. The amplitude and frequency of the external driving are $A=0.1$ and $\omega=0.1$, respectively. (b) The dependence of signal response on the frequency of the external driving force. The noise intensity is $D=0.2$.} %
\label{Fig:1}
\end{figure}
As Fig.~\ref{Fig:1} shows, the theoretical signal response of Eq.~(\ref{eq:LRTsinglebistable}) showcase a bell-shaped amplification curve as the noise intensity increases. Moreover, the signal response decreases monotonously in the form of the inverse-Sigmoid-curve as the driving frequency increases. Note that if the driving frequency is sufficiently high, i.e., $\omega>1.0$, the time window of the potential well variation is so small that it cannot afford the interwell motion of the bistable particle, as a result, the resonance-like signal response is disappeared. Consequently, a proper level of driving frequency is a pivotal factor for the emergence of SR. 

Furthermore, the theoretical signal response of Eq.~(\ref{eq:LRTsinglebistable}) agrees well with the theoretical response $\eta_{\textrm{single}}(\omega)=1/[D^{2}+D^{2}\omega^{2}\pi^{2}\exp(1/2D)/2]$ in ~\cite{Jung1991}, which verifies that the main processes and assumptions above are credible. We here considered the theoretical signal response of Eq.~(\ref{eq:LRTsinglebistable}) as the reference substance to highlight the role of other factors, like coupling and noise color, on signal response.

\subsection{Mean-field coupled oscillators within Gaussian white noise}
As the coupling among elements is indispensable in natural and artificial multiparticle systems, and the mean-field coupling has physical significance in modeling the muscle contraction and multimode solid lasers~\cite{Hu1996,Kang2008}, we here theoretically investigate the role of coupling on SR through the mean-field coupled overdamped bistable oscillators.
The dynamics of which driven by Gaussian white noise can be depicted by the stochastic differential equations 
\begin{equation}\label{eq:25}
\dot{x}_{i}=x_{i}-x_{i}^{3}+\frac{c}{N}\sum_{j=1}^{N}(x_{j}-x_{i})+A\sin(\omega t)+\xi_{i}(t),
\end{equation}
in which $i\in[1,N]$, the symbol $c$ is the strength of the mean-field coupling, the first and second moments of $\xi_{i}(t)$ are $\langle\xi_{i}(t)\rangle=0$ and $\langle\xi_{i}(t)\xi_{j}(t')\rangle =2D\delta_{ij}\delta(t-t')$, respectively. We further consider the spectral amplification factor of the mean oscillation, $X(t)=N^{-1}\sum\nolimits _{j=1}^{N}x_{j}(t)$, as
\begin{equation}\label{eq:Amplification factor}
\eta_{\textrm{meanfield}}=\frac{4}{A^{2}} \left| \langle e^{i\omega t}X(t)\rangle \right|^{2}, 
\end{equation}
to measure the collective signal response of coupled oscillators. We point out that the signal response in the present work is of ergodicity, namely, for the coupled bistable oscillators, the ensemble average of the signal response for the single particle is the same as the signal response of the average dynamics of the mean-field coupled oscillators~\cite{Tessone2006,Liu2023}. 

Additionally,  Eq.~(\ref{eq:25}) can be further recast as
\begin{equation}\label{eq:overdampedmeanfield}
\dot{x}=x-x^{3}+c(X(t)-x)+A\sin(\omega t)+\xi(t). 
\end{equation}
Since all the oscillators are identical, we can neglect the subscript. Moreover, as Lucarini revealed in reference~\cite{Lucarini2020}, one can expand the probability density function as $W(x,t)=W_{\textrm{st}}(x)+\epsilon W_{\textrm{1}}(x,t)+O(\epsilon^{2} )$, in which the symbol $\epsilon$ is a small quantity. Since $X(t)=\int xW(x,t)dx$, $X_{\textrm{0}}=\int xW_{\textrm{st}}(x)dx$, and $X_{\textrm{1}} (t)=\int xW_{\textrm{1}} (x,t)dx$, etc., we can obtain the average motion as $X(t)=X_{\textrm{0}}(t)+X_{\textrm{1}} (t)+O(\epsilon^{2} )$. However, in the present work, under LRT, we neglect the high-order terms and only retain the first two terms as $X(t)=X_{\textrm{0}}(t)+X_{\textrm{1}} (t)$, in which the term $X_{\textrm{0}}(t)$ is a constant representing the behavior at steady state and the symbol $X_{\textrm{1}}(t)$ depicts the oscillation under the influence of external signal. Bring $X(t)$ into Eq.~(\ref{eq:overdampedmeanfield}), the Fokker-Planck equation for Eq.~(\ref{eq:overdampedmeanfield}) can be depicted as 
\begin{eqnarray}\label{eq:27}
\frac{\partial W(x,t)}{\partial t}&=&-\frac{\partial}{\partial x}\left\{\left[(1-c)x-x^{3}+cX_{0}\right]W(x,t)\right\}\nonumber\\
&+&D\frac{\partial^{2}}{\partial x^{2}}W(x,t)-\left[cX_{1}+A\sin(\omega t)\right]\frac{\partial}{\partial x}W(x,t)\nonumber\\
&=&L_{0}W(x,t)+L_{1}W(x,t),
\end{eqnarray}
where $L_{0}=-\partial\left[(1-c)x-x^{3}+cX_{0}\right]/\partial x$ and $L_{1}=-\partial\left[cX_{1}+A\sin(\omega t)\right]/\partial x$.

Following the vein of the analytical processes of the single bistable oscillator, the mean oscillation caused by the external driving can be depicted by
\begin{equation}\label{eq:30}
X_{1}=\int_{-\infty}^{t}\chi(t-t')\left[cX_{1}+A\sin(\omega t)\right]dt'.
\end{equation}
It is worthy pointing out that the potential function of Eq.~(\ref{eq:overdampedmeanfield}) for a vanished driving is no longer symmetrical due to the mean-field coupling. Nevertheless, under the weak mean-field coupling assumption, we can still consider approximatively $\lambda$ as the escape rate of the mean-filed coupled oscillator. As a result, we can obtain the correlation function
\begin{equation}\label{eq:Meanfieldcorrelation}
\langle x(t)x(t')\rangle_{0}=\langle x(t)^{2} \rangle_{0} e^{-\lambda t'},
\end{equation}
and thereby, the response function can be further written as 
\begin{equation}\label{eq:20}
\chi(t-t')=\lambda\frac{\langle x(t)^{2}\rangle_{0}}{D}e^{-\lambda t'}.   
\end{equation}
After Fourier transformation of Eq.~(\ref{eq:30}), we have 
\begin{equation}\label{eq:31}
\widetilde{X}_{1}(\omega)=\frac{\widetilde{\chi}(\omega)}{1-c\widetilde{\chi}(\omega)}\widetilde{F}_{\textrm{ext}}(\omega)=\widetilde{\chi}_{t}(\omega)\widetilde{F}_{\textrm{ext}}(\omega),
\end{equation}
where
\begin{equation}\label{eq:32}
\widetilde{\chi}_{t}(\omega)=\lambda\frac{\langle x(t)^{2}\rangle_{0}}{D}\frac{-1}{i\omega+\left(c\frac{\langle x(t)^{2}\rangle_{0}}{D}-1\right)\lambda}
\end{equation}
Finally, the signal response of Eq.~(\ref{eq:25}) is
\begin{equation}\label{eq:33}
\eta_{\textrm{meanfield}}(\omega)=\left(\lambda\frac{\langle x(t)^{2}\rangle_{0}}{D}\right)^{2}\frac{1}{\omega^{2}+\left(c\frac{\langle x(t)^{2}\rangle_{0}}{D}-1\right)^{2}\lambda^{2}}.
\end{equation}

As the blue dashed line in Fig.~\ref{Fig:2} (a) shows, a clear resonance-like signal response curve can be viewed when the noise intensity is gradually increased. Besides, as the driving frequency increases, the signal response decreases monotonously, see Fig.~\ref{Fig:2} (b). Furthermore, compared to the signal response of the single overdamped particle system, the mean-field coupling can significantly enhance the magnitude of SR especially for a low-frequency signal. These theoretical results are compatible with the numerical or the semi-analytical conclusions in the mean-field coupled overdamped bistable oscillators with respect to the Gaussian white noise~\cite{Kang2008}. 

Since the mean-field coupled bistable oscillators have been considered as a testbed for investigating the role of coupling on SR in the last three decades, some acclaimed theoretical signal response results have been proposed. For instance, Jung and coworkers deduced the theoretical spectral power amplification at the driving frequency by the two-state master equation~\cite{Jung1992}. Pikovsky \emph{et al}. proposed the signal response formula based on the Gaussian approximation and the slaving principle~\cite{Pikovsky2002}. Neiman \emph{et al}. gave the signal response formula of the ensemble of $N$ uncoupled bistable elements~\cite{Neiman1997}. Based on the formalism of the Graham's quasi-potential and master equation, Lucarini deduced both the spectrum amplification and signal-to-noise ratio of $N$-dimensional nonequilibrium systems~\cite{Lucarini2019}. Despite these findings, to the best of our knowledge, a concise theoretical interpretation of the mean-field coupled bistable oscillators through the LRT viewpoint is still lacking. The result of Eq.~(\ref{eq:33}) enriches the theoretical understanding of SR in this case.  

\subsection{Single oscillator under OU noise}
Notwithstanding the Gaussian white noise is widely used in SR research, it is
inaccessible to an experimentalist or the physical world because the generation of white noise requires an infinite amount of power~\cite{Hanggi1993,Hanggi1994}. On the contrary, the colored noise concerning a finite correlation time is encountered in a wide variety of natural and man-made systems and has been recognized as an emblematic feature of complexity~\cite{Gilden1995}. One key issue concerning colored noise is whether such a pervasive type of noise outperforms white noise in weak signal response~\cite{Nozaki1999}. We here theoretically explore this question. Moreover, to compare carefully the signal responses between the colored noise and white noise situations, we first focus on the case of a single particle driven by both the classical OU noise and periodical signal, the dynamics of which can be represented by the stochastic differential equation
\begin{equation}\label{eq:OU}
\dot{x}=x-x^{3}+A\sin(\omega t)+\zeta(t), 
\end{equation}
in which the first and second moments of $\zeta(t)$ are $\langle\zeta(t)\rangle=0$ and $\langle \zeta(t)\zeta(t')\rangle=\frac{D}{\tau}e^{-\frac{|t-t'|}{\tau}}$, respectively. The symbol $\tau$ is the correlation time or noise color. The power spectrum, $S_{\textrm{OU}}=\int_{-\infty}^{\infty}\langle\zeta(t)\zeta(t')\rangle e^{-i\omega t}dt=2D/(1+\tau^{2}\omega^{2})$, is bounded, which characterizes a random process with finite correlation time. As a consequence, the OU noise is conventionally considered as one typical colored noise. Moreover, the non-Markovian one-dimensional evolution of Eq.~(\ref{eq:OU}) can be recast into the Markovian bidimensional process $(x,y)$, which can be further depicted by the Langevin equation 
\begin{equation}\label{eq:Bi-dimensional}
\dot{x}=x-x^{3}+A\sin(\omega t)+y;\dot{y}=-\frac{y}{\tau} +\frac{\xi(t)}{\tau},
\end{equation}
where $\xi(t)$ are Gaussian random variables with $\langle\xi(t)\rangle=0$ and $\langle \xi(t)\xi(t') \rangle=2D\delta(t-t')$. Based on the effective diffusion approximation where the Fokker-Planck of the OU noise case can be retained as the white noise form but the diffusion strength is substituted by the effective diffusion~\cite{Hanggi1993}
\begin{equation}\label{eq:34}
D_{\textrm{eff}}\to\frac{D}{1-\tau(1-3\left\langle x^{2}\right\rangle_{0})}.
\end{equation}
Therefore, the Fokker-Planck equation for single overdamped bistable oscillator driven by OU noise can be recast as 
\begin{eqnarray}\label{eq:35}
\frac{\partial W(x,t)}{\partial t}&=&-\frac{\partial}{\partial x}\left[(x-x^{3})W(x,t)\right]-A\sin(\omega t)\frac{\partial}{\partial x}W(x,t)\nonumber\\
&+&\frac{D}{1-\tau(1-3\left\langle x^{2}\right\rangle_{0})}\frac{\partial^{2}}{\partial x^{2}}W(x,t).
\end{eqnarray}
Previously, H\"{a}nggi et al. have shown that the response function of a system with OU noise should satisfy the following correlation~\cite{Hanggi1993}
\begin{equation}\label{eq:36}
\chi(t-t')=-\frac{1}{D}\frac{\mathrm{d}}{\mathrm{d}t'}\left\langle x(t)\psi\left[x(t')\right]\right\rangle_{0},
\end{equation}
where $\psi\left[x(t)\right]=x(t)+\tau\left[x(t)-x^{3}(t)\right]$. Moreover, we can still assume that the correlation function has a relationship in time as 
\begin{equation}\label{eq:37}
\left\langle x(t)\psi\left[x(t')\right]\right\rangle_{0}=\left\langle x(t)\psi\left[x(t)\right]\right\rangle_{0} e^{-\lambda_{\textrm{OU}}t'},
\end{equation}
in which the Kramers escape rate under the OU noise~\cite{Hanggi1993} is   
\begin{equation}\label{eq:38}
\lambda_{\textrm{OU}}=\frac{1}{\sqrt{2}\pi}(1-\frac{3}{2}\tau)e^{-\frac{1}{4D}}.
\end{equation}
Inserting Eq.~(\ref{eq:37}) and Eq.~(\ref{eq:38}) into Eq.~(\ref{eq:36}), then we have 
\begin{equation}\label{eq:39}
\chi(t-t')=\lambda_{\textrm{OU}}\frac{\left\langle x(t)\psi\left[x(t)\right]\right\rangle_{0}}{D}e^{-\lambda_{\textrm{OU}}t'}.
\end{equation}
Specifically, the correlation function $\left\langle x(t)\psi\left[x(t)\right]\right\rangle_{0}$ can be recast as,
\begin{equation}\label{eq:40}
\left\langle x(t)\psi\left[x(t)\right]\right\rangle_{0}=\left\langle x^{2}(t)\right\rangle_{0}+\tau\left\langle x(t)F(t)\right\rangle_{0},
\end{equation}
where $F(t)=x-x^{3}$ can be viewed as the force on the particle with no external drive, then the term $\left\langle x(t)F(t)\right\rangle_{0}$ is actually the power of the noise. For a small $\tau$, we have 
\begin{equation}\label{eq:41}
\left\langle x(t)F(t)\right\rangle_{0}=\lim_{\tau\to 0}\int_{-\infty}^{\infty}\left\langle\zeta(t)\zeta(t')\right\rangle e^{-i\omega t}dt=D,\nonumber
\end{equation}
As a result, the response function of Eq.~(\ref{eq:OU}) is actually
\begin{equation}\label{eq:42}
\chi(t-t')=\lambda_{\textrm{OU}}\frac{\left\langle x^{2}(t)\right\rangle_{0}+\tau D}{D}e^{-\lambda_{\textrm{OU}}t'}.
\end{equation}
After Fourier transform, Eq.~(\ref{eq:42}) can be written as 
\begin{equation}\label{eq:43}
\widetilde{\chi}(\omega)=\lambda_{\textrm{OU}}\frac{\left\langle x^{2}(t)\right\rangle_{0}+\tau D}{D}\left(\frac{-1}{i\omega-\lambda_{\textrm{OU}}}\right),
\end{equation}
and thereby the signal response can be obtained as
\begin{equation}\label{eq:OUsingleoverdampedoscillator}
\eta_{\textrm{single}}(\omega)=\left[\lambda_{\textrm{OU}}\frac{\left\langle x^{2}(t)\right\rangle_{0}+\tau D}{D}\right]^{2}\left(\frac{1}{\omega^{2}+\lambda^{2}_{\textrm{OU}}}\right) .
\end{equation}
\begin{figure}
\centering
\includegraphics[width=1.0\columnwidth]{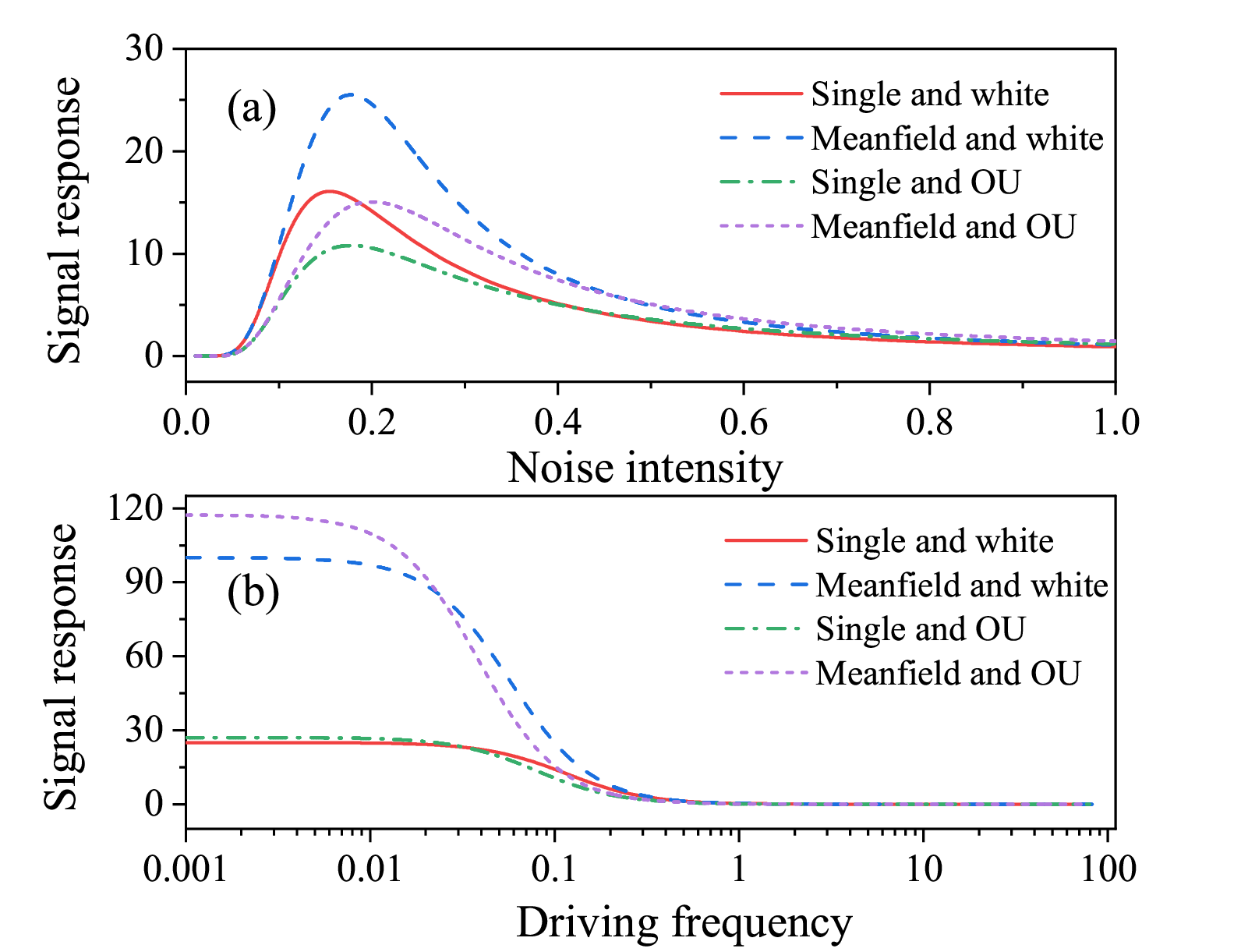}
\caption{ Stochastic resonance in overdamped bistable systems. (a) Signal response versus the noise intensity for a single particle and the mean-field coupled system under the influence of the Gaussian white noise and the OU noise, respectively. The amplitude and frequency of the external signal are both $0.1$. The strength of mean-field coupling is $c=0.1$, the correlation time is $\tau=0.2$. (b) Relationship between signal response and the external driving frequency for distinct situations. The noise intensity is $D=0.2$.} 
\label{Fig:2}
\end{figure}
As the green dot-dashed line in Fig.~\ref{Fig:2} (a) shows, the resonant signal response can be also observed through Eq.~(\ref{eq:OUsingleoverdampedoscillator}). Furthermore, compared to the white noise situation, the optimal signal response and the optimal noise intensity are reduced and increased, respectively. These results are in agreement with the conventional conclusion that noise color weakens the magnitude of SR in the overdamped case. 
Additionally, with the increase of the driving frequency, the two signal response curves corresponding to the OU and white noise situations showcase a intersection, see in Fig.~\ref{Fig:2} (b). In detail, if the driving frequency is less than the critical frequency, like $\omega=0.02$, the signal response of the single particle driven by the OU noise outperforms the white noise counterpart. On the contrary, if the driving frequency is larger than the critical frequency, the white noise case is advantaged.   

Such a crossing phenomenon can be also explained from the theoretical results of Eq.~(\ref{eq:LRTsinglebistable}) and Eq.~(\ref{eq:OUsingleoverdampedoscillator}). For a low-frequency signal, the optimal noise intensity of the Gaussian white noise case approaches to a smaller level based on the Kramers escape rate, the given noise intensity $D=0.2$ is thus supra-optimal for this case. Nevertheless, due to the signal response advantages in the strong noise region of the OU noise situation, the signal response of the OU noise case is superior to the white noise counterpart for a low-frequency signal. 

Note that, Nozaki and collaborators put forward the conjecture that colored noise may have functional significance compared to white noise in weak signal response~\cite{Nozaki1999}. Moreover, for a fixed driving frequency, they verified that the signal response evoked by colored noise of excitable neurons is superior to that by white noise for a sufficient strong noise. Nevertheless, how the driving frequency influences their conclusion is still unknown. The result of Eq.~(\ref{eq:OUsingleoverdampedoscillator}) shows explicitly that the signal response of the OU noise case can be better than that of the Gaussian white noise, the result of Eq.~(\ref{eq:LRTsinglebistable}), if the driving frequency is sufficiently low. These results complement the conclusion in reference~\cite{Nozaki1999} concerning the functional significance of colored noise in signal response.

\subsection{Mean-field coupled oscillators under OU noise} 
Next, we concentrate on exploring the signal response of the mean-filed coupled oscillators driven by the OU noise and weak stimulus. The dynamical evolution of such a case can be represented by 
\begin{equation}\label{eq:MeanfieldOU}
\dot{x}_{i}=x_{i}-x_{i}^{3}+\frac{c}{N}\sum_{j=1}^{N}(x_{j}-x_{i})+A\sin(\omega t)+\zeta_{i}(t),
\end{equation}
in which the first-order and second-order moments of the OU noise $\zeta_{i}(t)$ are respectively 
\begin{equation}\label{eq:44}
\langle\zeta_{i}(t)\rangle=0,\langle\zeta_{i}(t)\zeta_{j}(t')\rangle=\delta_{ij}\frac{D}{\tau}e^{-\frac{|t-t'|}{\tau}}.
\end{equation}
Similar to the deducing processes in the white noise counterpart, Eq.~(\ref{eq:MeanfieldOU}) can be reduced to a mean-field equation and the response function is  
\begin{equation}\label{eq:45}
\widetilde{\chi}_{t}(\omega)=\frac{\widetilde{\chi}(\omega)}{1-c\widetilde{\chi}(\omega)}=\frac{-\lambda_{\textrm{OU}}\frac{\left\langle x^{2}(t)\right\rangle_{0}+\tau D}{D}}{i\omega+\left(c\frac{\left\langle x^{2}(t)\right\rangle_{0}+\tau D}{D}-1\right)\lambda_{\textrm{OU}}},
\end{equation}
the resulting signal response is
\begin{align}\label{eq:46}
\eta_{\textrm{meanfield}}(\omega)=\left(\lambda_{\textrm{OU}}\frac{\alpha}{D}\right)^{2}\frac{1}{\omega^{2}+\left(c\frac{\alpha}{D}-1\right)^{2}\lambda^{2}_{\textrm{OU}}},
\end{align}
where $\alpha=\left\langle x^{2}(t)\right\rangle_{0}+\tau D$.

As shown in Fig.~\ref{Fig:2} (a), we can find that the theoretical signal response of the mean-field coupled oscillators of Eq.~(\ref{eq:46}) captures the main resonance character and is larger than that of the single oscillator case. However, compared to the mean-field coupled counterpart disturbed by the Gaussian white noise, the SR magnitude evoked by the OU noise is reduced. 
As Fig.~\ref{Fig:2} (b) shows, besides the monotonically decreasing signal response, an obvious crossing signal response curves between the Gaussian white and the OU noise cases can be also viewed with the rise of the driving frequency.
The reason for such a phenomenon is same as the single particle situations.

It is worthy pointing out that although H\"{a}nggi and coworkers have systematically investigated the influence of diverse types of colored noise on SR in a single bistable case~\cite{Hanggi1993}, how colored noise influences the collective signal response of the mean-filed coupled bistable oscillators still needs to be theoretically revealed. The result of Eq.~(\ref{eq:46}) enriches the theoretical understanding of such a case. Additionally, resembling the conclusion in case of a single particle driven by the OU noise, the theoretical description of Eq.~(\ref{eq:46}) explicitly verifies that colored noise has functional significance in signal response, especially for a low frequency driving. 

\begin{figure}
\centering
\includegraphics[width=1.0\columnwidth]{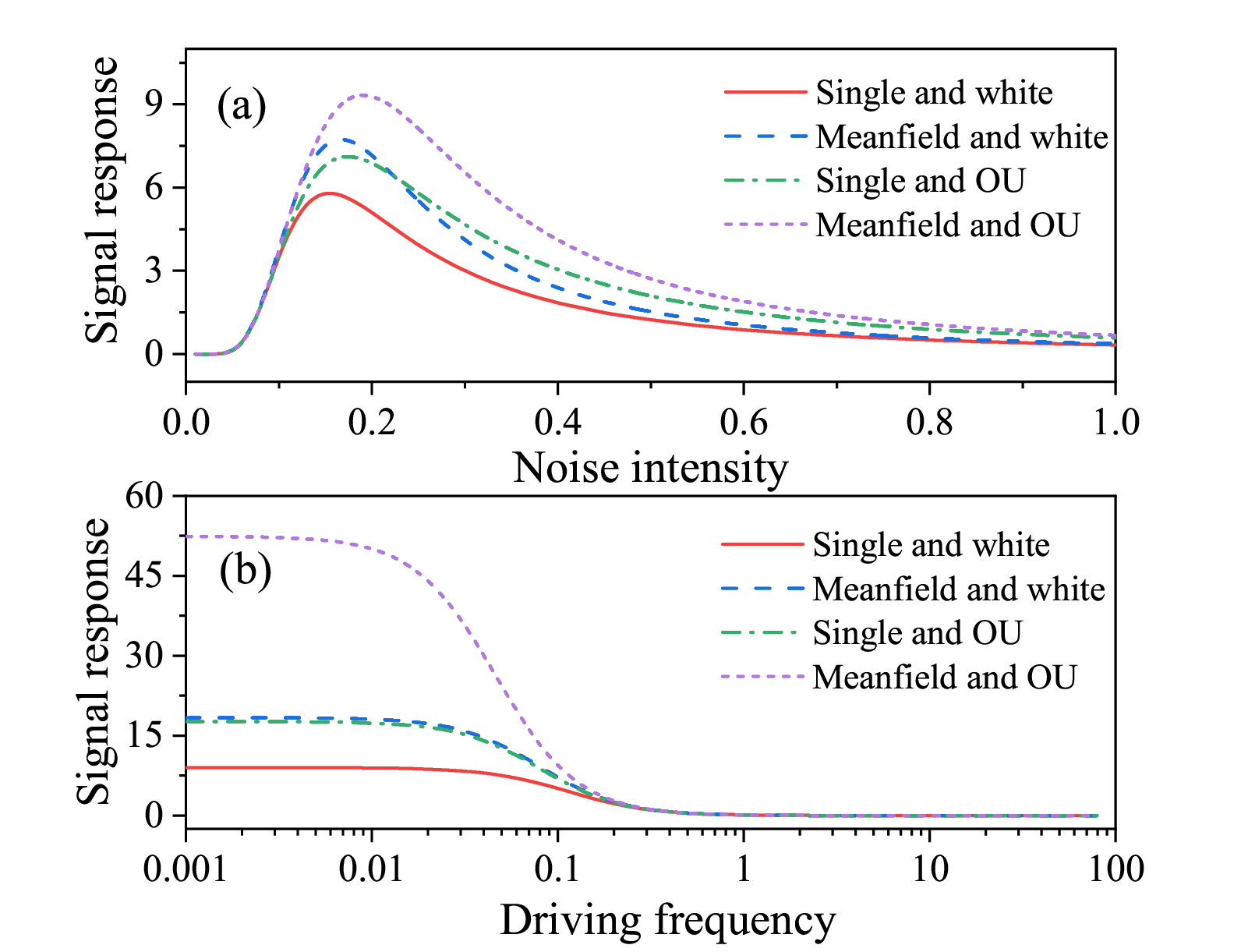}
\caption{ Stochastic resonance in underdamped Duffing systems. (a) Signal response versus the noise intensity for a single particle and the mean-field coupled oscillators under the influence of Gaussian white noise and the OU noise, respectively. The amplitude and frequency of the external signal are both $0.1$. The damping coefficient is $\gamma=0.6$. The strength of mean-field coupling is $c=0.1$, the correlation time is $\tau=0.2$. (b) Dependence of signal response on the external driving frequency for diverse cases. The noise intensity is $D=0.2$.} 
\label{Fig:3}
\end{figure} 
 
\section{SR For Underdamped Duffing model}\label{Sec2B}

\subsection{Single Duffing oscillator and Gaussian white noise }
Since the Duffing equation is widely acclaimed in fields ranging from mechanical and engineering to biophysics~\cite{Liu2019,Peters2021,Liang2021}, and showcases the bistable characteristic as well. We continue to investigate the theoretical signal response in such models. 
Additionally, resembling the spirit in the overdamped case, we here first consider the signal response of a single Duffing oscillator and its dynamics can be written as   
\begin{equation}\label{eq:SingleDuffing}
\ddot{x}+\gamma\dot{x}=x-x^{3}+A\sin(\omega t)+\xi(t),  
\end{equation}
in which $x$ is also the state variable, the symbol $\gamma$ is the damping coefficient.
Conventionally, Eq.~(\ref{eq:SingleDuffing}) can be further recast into the two-dimensional Langevin equations as
\begin{equation}\label{eq:LangevinsingleDuffing}
\dot{x}=v;\dot{v}=x-x^{3}-\gamma v+A\sin(\omega t)+\xi(t), 
\end{equation}
where $v(t)$ can be viewed as the velocity, and thereby the corresponding Fokker-Planck equation can be further read as 
\begin{eqnarray}\label{eq:FPduffingsingle}
&&\frac{\partial W(x,v,t)}{\partial t}=-v\frac{\partial}{\partial x}W(x,v,t)+D\gamma\frac{\partial^{2}}{\partial v^{2}}W(x,v,t)\nonumber\\
&-&\frac{\partial}{\partial v}\left\{\left[x-x^{3}-\gamma v+A\sin(\omega t)\right]W(x,v,t)\right\}\nonumber\\
&=&-v\frac{\partial}{\partial x}W(x,v,t)-\frac{\partial}{\partial v}\left\{\left[x-x^{3}-\gamma v\right]W(x,v,t)\right\}\nonumber\\
&+&D\gamma\frac{\partial^{2}}{\partial v^{2}}\left\{W(x,v,t)\right\}-A\sin(\omega t)\frac{\partial}{\partial v}W(x,v,t)\nonumber\\
&=&L_{0}W(x,v,t)+L_{1}W(x,v,t).
\end{eqnarray}
The steady state probability density of Eq.~(\ref{eq:FPduffingsingle}) is 
\begin{eqnarray}\label{eq:Steadystatedensity}
&&W_{\textrm{st}}(x,v)=Z\exp\nonumber\\
&&\left(-\left[\ln D^{(2)}(x)-\int\int\frac{D^{(1)}(x',v')}{D^{(2)}(x')}dx'dv'\right]\right)\nonumber\\
&=&Z\exp\left(-\ln(\gamma D)+\frac{1}{\gamma D}\int\int(x-x^{3}-\gamma v)dxdv\right)\nonumber\\
&=&Z\gamma D\exp\left(\frac{1}{\gamma D}\left(\frac{1}{2}x^{2}-\frac{1}{4}x^{4}-\frac{\gamma}{2}v^{2}\right)\right),
\end{eqnarray}
in which the two symbols $D^{(1)}=x-x^{3}-\gamma v+A\sin(\omega t)$ and $D^{(2)}=D\gamma$ are the drift and diffusion terms, respectively. Still, based on the linear response theory, we know that 
\begin{equation}
W(x,v,t)=W_{\textrm{st}}(x,v)+W_{\textrm{ext}}(x,v,t),  
\end{equation}
and through the path integral method~\cite{West1988}, we have the 
\begin{equation}
W_{\textrm{ext}}(x,v,t)=\int_{-\infty}^{t}e^{L_{0}t'}L_{1}(t')W_{\textrm{st}}(x,v)dt' ,  
\end{equation}
Considering the external perturbation operator as
\begin{equation}
L_{1}=-A\sin(\omega t)\frac{\partial}{\partial v}=F_{\textrm{ext}}(t)\tilde{L}_{1}.   
\end{equation}
where $F_{\textrm{ext}}(t)$ and $\tilde{L}_{1}$ are the external force and perturbation operator, respectively. Thus, the eigenvalue equation is
\begin{equation}
\tilde{L}_{1}W_{\textrm{st}}(x,v)=-\frac{\partial}{\partial v}W_{\textrm{st}}(v)=\frac{\gamma v}{D}W_{\textrm{st}}(x,v) ,
\end{equation}
Since the mean of oscillation under the external perturbation is  
\begin{equation}\label{eq:MeansingleDuffing1}
\left\langle x_{1}\right\rangle=\int\int x(t)W_{\textrm{ext}}(x,v)dxdv,    
\end{equation}
inserting $W_{\textrm{ext}}(x,v,t)$ into Eq.~(\ref{eq:MeansingleDuffing1}), we obtain 
\begin{eqnarray}\label{eq:CalculationsingleDuffing}
\left\langle x_{1}\right\rangle=\int\int xW_{\textrm{ext}}(x,v)dxdv=\int^{t}_{-\infty}R(t-t')F_{\textrm{ext}}(t')dt'.
\end{eqnarray}
As a result, the response function can be written as 
\begin{equation}
R(t-t')=\frac{\gamma}{D}\int_{x}\int_{v}xve^{L_{0}(t-t')}W_{\textrm{st}}(x,v)dxdv.    
\end{equation}
The deducing procedures from the mean oscillation to the response function can be viewed as the response theory for the measurement to response theory for observables.

If we consider the transition probability from the state $(x,v,t)$ to $(x',v',t')$ as $P(x,v,t|x',v',t')$,  then we can have the formulas
\begin{equation}
P(x,v,t|x',v',t')=e^{L_{0}(t-t')}\delta(x-x')\delta(v-v'),
\end{equation}  
and further
\begin{equation}
P(x,v,t|x',v',t')P_{0}(x,v)=e^{L_{0}(t-t')}P_{0}(x',v').    
\end{equation}
The derivative of the transition probability with respect to time is 
\begin{eqnarray}
\frac{\mathrm{d}}{\mathrm{d}t'}P(x,v,t|x',v',t')&=&-e^{L_{0}(t-t')}L_{0}(t-t')\delta(x-x')\delta(v-v'). \nonumber    
\end{eqnarray}
As a consequence, the response function can be expressed as 
\begin{eqnarray}\label{eq:resonsefunctionofsingleDuffing12}   
R(t-t')&=&\frac{\gamma}{D}\int_{x}\int_{v}xve^{L_{0}(t-t')}W_{\textrm{st}}(x,v)dxdv\nonumber\\
&=&\frac{\gamma}{D}\frac{\mathrm{d}}{\mathrm{d}t'}\left\langle x(t)x(t')\right\rangle_{0}.
\end{eqnarray}
Noticeably, the Eq.~(\ref{eq:resonsefunctionofsingleDuffing12}) suggests that the response function of the underdamped situation has a same form as the overdamped counterpart. Furthermore, the potential function of the underdamped case is still bistable, we thus consider the correlation function has a same modality as overdameped case as 
\begin{equation}
\left\langle x(t)x(t')\right\rangle_{0}=\left\langle x(t)^{2}\right\rangle_{0} e^{-\lambda t'},  
\end{equation}
consequently, the response function is 
\begin{equation}
R(t-t')=\frac{\gamma}{D}\frac{\mathrm{d}}{\mathrm{d}t'}\left\langle x(t)x(t')\right\rangle_{0}=-\lambda\frac{\gamma}{D}\langle x(t)^{2}\rangle_{0} e^{-\lambda t'}.
\end{equation}
After performing a Fourier transform, we have 
\begin{equation}
R(\omega)=-\frac{\lambda\gamma\langle x(t)^{2}\rangle_{0}}{D}\frac{1}{i\omega-\lambda},    
\end{equation}
and the signal response can be written as 
\begin{equation}\label{eq:SingleDuffingwhitenoise}
\eta_{\textrm{single}}(\omega)=\left[\frac{\lambda\gamma\langle x(t)^{2}\rangle_{0}}{D}\right]^{2}\frac{1}{\omega^{2}+\lambda^{2}}.   
\end{equation}
In Fig.~\ref{Fig:3}, we can find a clear bell-shaped signal response profile, shown in red solid line, as the noise intensity increases. Besides, with the increase of the external driving frequency, the signal response decreases as the inverse-Sigmoid-curve as shown in the overdamped situations. This result demonstrates that a proper frequency is also a prerequisite for the occurrence of SR in underdamped Duffing systems.

\subsection{Mean-field coupled Duffing oscillators and white noise }
Furthermore, for the case of the mean-field coupled Duffing elemetns driven by both the Gaussian white noise and weak external signal, the dynamics can be written as 
\begin{eqnarray}\label{eq:MeanfieldDuffing2}
\dot{x_{i}}&=&v_{i}\nonumber\\
\dot{v_{i}}&=&(1-c)x_{i}-x_{i}^{3}-\gamma v_{i}+cX+A\sin(\omega t)+\xi_{i}(t).
\end{eqnarray}
Since all the oscillators are identical, we neglect the subscript of oscillators and thereby the corresponding Fokker-Planck equation of Eq.~(\ref{eq:MeanfieldDuffing2}) is
\begin{eqnarray}
\frac{\partial W(x,v,t)}{\partial t}&=&-v\frac{\partial}{\partial x}W(x,v,t)+D\gamma\frac{\partial^{2}}{\partial v^{2}}W(x,v,t)\nonumber\\
&-&\frac{\partial}{\partial v}[\beta+cX^{1}+A\sin(\omega t)]W(x,v,t)\nonumber\\
&=&-v\frac{\partial}{\partial x}W(x,v,t)+\gamma D\frac{\partial^{2}}{\partial v^{2}}\left\{W(x,t)\right\}\nonumber\\
&-&\frac{\partial}{\partial v}\left\{\beta W(x,v,t)\right\}\nonumber\\
&-&\left[cX^{1}+A\sin(\omega t)\right]\frac{\partial}{\partial v}W(x,t)\nonumber\\
&=&L_{0}W(x,t)+L_{1}W(x,t),
\end{eqnarray}
in which $\beta=(1-c)x-x^{3}-\gamma v+cX^{0}$. The steady-state probability density is
\begin{eqnarray}\label{eq:steady-stateprobabilitydensity}
W_{\textrm{st}}(x,v)&=&Z\exp\left(-\left[\ln D^{(2)}(x)-\int\int\frac{D^{(1)}(x',v')}{D^{(2)}(x')}dx'dv'\right]\right)\nonumber\\
&=&Z\exp\left(-\ln(\gamma D)+\frac{1}{\gamma D}\int\int\beta dxdv\right)\nonumber\\
&=&Z\gamma D\exp\left(\frac{1}{\gamma D}\left[\frac{1-c}{2}x^{2}-\frac{1}{4}x^{4}-\frac{\gamma}{2}v^{2}+cxX^{0}\right]\right).
\end{eqnarray}
Similarly, through the path integral method, we can obtain 
\begin{equation}
W_{\textrm{ext}}(x,v,t)=\int_{-\infty}^{t}e^{L_{0}t'}L_{1}(t')W_{\textrm{st}}(x,v)dt' .   
\end{equation}
The external perturbation operator $L_{1}$ can be recast as 
\begin{equation}
L_{1}=-\left[cX^{1}+A\sin(\omega t)\right]\frac{\partial}{\partial v}=F_{\textrm{ext}}(t)\tilde{L}_{1}.    
\end{equation}
The resulting eigenvalue equation is 
\begin{equation}
\tilde{L}_{1}W_{\textrm{st}}(x,v)=-\frac{\partial}{\partial v}W_{\textrm{st}}(v)=\frac{\gamma v}{D} W_{\textrm{st}}(x,v).    
\end{equation}
Additionally, since the mean oscillation of the Duffing oscillators is
\begin{equation}
X^{1}(t)=\int\int x(t)W_{\textrm{ext}}(x,v)dxdv,   
\end{equation}
we can obtain 
\begin{eqnarray}\label{eq:meanDuffing3}
X^{1}(t)=\int\int xW_{\textrm{ext}}(x,v)dxdv=\int^{t}_{-\infty}R(t-t')F_{\textrm{ext}}(t')dt'.
\end{eqnarray}
Succinctly, the Eq.~(\ref{eq:meanDuffing3}) can be recast as 
\begin{equation}\label{eq:meanDuffing12}
X^{1}(t)=\int^{t}_{-\infty}R(t-t')\left[cX^{1}(t)+A\sin(\omega t)\right]dt'.    
\end{equation}

Performing a Fourier transform on Eq.~(\ref{eq:meanDuffing12}), then we have 
\begin{equation}
\tilde{X}_{1}(\omega)=\chi(\omega)\left(c\tilde{X}_{1}(\omega)+F(\omega)\right),    
\end{equation}
and the resulting relation
\begin{equation}
\tilde{X}_{1}(\omega)=\frac{\chi(\omega)}{1-c\chi(\omega)}F(\omega)=\tilde{\chi}(\omega)F(\omega).    
\end{equation}
Resembling the situation of the single Duffing element, we can obtain the response function as
\begin{eqnarray}
R(t-t')&=&\frac{1}{D}\int_{x}\int_{v}xve^{L_{0}(t-t')}W_{\textrm{st}}(x,v)dxdv\nonumber\\
&=&\frac{1}{D}\gamma\frac{\mathrm{d}}{\mathrm{d}t'}\left\langle x(t)x(t')\right\rangle_{0}\nonumber\\
&=&-\frac{1}{D}\gamma\lambda\gamma \langle x(t)^{2}\rangle_{0} e^{-\lambda t'}.
\end{eqnarray}
Furthermore, after the Fourier transform, the response function is recast as
\begin{equation}
\chi(\omega)=-\frac{1}{D}\lambda\gamma\langle x(t)^{2}\rangle_{0} \frac{1}{i\omega-\lambda},    
\end{equation}
which leads to the form
\begin{equation}
\tilde{\chi}(\omega)=\frac{\chi(\omega)}{1-c\chi(\omega)}=\frac{\frac{1}{D} \lambda\gamma\langle x(t)^{2}\rangle_{0} }{\left[1-\frac{c\gamma\langle x(t)^{2}\rangle_{0}}{D}\right]\lambda-i\omega}.
\end{equation}
Of particular note is that an  analogous yet more general susceptibility expression $\tilde{\chi}(\omega)=\sum_{k}g_{k}(i\omega-\lambda_{k})^{-1}$, in which $g_{k}$ and $\lambda_{k}$ are coefficients and eigenvalues, can be found in~\cite{Lucarini2020,Zagli2024}. 

Finally, the signal response of the mean-field coupled Duffing oscillators driven by the Gaussian white noise can be written as 
\begin{equation}\label{eq:finalresponse1}
\eta_{\textrm{meanfield}}(\omega)=\left[\frac{\lambda\gamma\langle x(t)^{2}\rangle_{0}}{D}\right]^{2}\frac{1}{\left[1-\frac{c\gamma \langle x(t)^{2}\rangle_{0}^{2}}{D}\right]\lambda^{2}+\omega^{2}}.
\end{equation}

As Fig.~\ref{Fig:3} shows, the resonance-like signal response can be amplified by the mean-filed coupling, and such a behavior is significant especially for the low-frequency signal.  

Despite the theoretical description for SR evoked by the Gaussian white noise in the underdamped environment are more complicated than that for the overdamped one, many outstanding results and methods have been consecutively revealed. For instance, for a single Duffing element, Gammaitoni \emph{et al}. theoretically investigated the relaxation properties of an underdamped bistable oscillator under the adiabatic approximation. Stocks \emph{et al}. deduced the theoretical signal response of SR in a monostable Duffing oscillator based on LRT. Kang and coworkers semi-analytically explored SR in an underdamped bistable Duffing oscillator by the method of moments~\cite{Kang2003}. For the coupled Duffing systems, Kang \emph{et al}. further explored signal response and bifurcation of the globally coupled underdamped Duffing oscillators by using the same moments method in references ~\cite{Kang2003,Liu2019}. Moreover, Lucarini investigated the SR evoked by the white noise in coupled oscillators displaying chaotic behavior in the framework of two-state theory~\cite{Lucarini2019}.
Nevertheless, like in the situation of overdamped oscillators, a compendious LRT description for the mean-filed coupled bistable Duffing systems is still lacking. The results of Eqs.~(\ref{eq:SingleDuffingwhitenoise}) and~(\ref{eq:finalresponse1}) contribute to the theoretical understanding of SR in Duffing systems.
\subsection{ Single Duffing oscillator or mean-filed coupled Duffing elements under OU noise }
Likewise the spirit in the overdamped bistable oscillators, we here continue to consider the single Duffing oscillator under the influence of the OU noise, the dynamics of which can be depicted by   
\begin{equation}
\ddot{x}+\gamma\dot{x}=x-x^{3}+A\sin(\omega t)+\zeta(t),  
\end{equation}
the resulting three-dimensional Langevin equations are
\begin{eqnarray}\label{eq:LangevinsingleDuffing}
\dot{x}=v;\dot{v}=x-x^{3}-\gamma v+A\sin(\omega t)+y;\dot{y}=-\frac{1}{\tau}y+\frac{1}{\tau}\xi(t).\nonumber
\end{eqnarray}
Furthermore, the corresponding effective Fokker-Planck equation can be read as 
\begin{eqnarray}\label{eq:FPduffing}
&&\frac{\partial W(x,v,t)}{\partial t}=-v\frac{\partial}{\partial x}W(x,v,t)\nonumber\\
&-&\frac{\partial}{\partial v}\left\{\left[x-x^{3}-\gamma v+A\sin(\omega t)\right]W(x,v,t)\right\}\nonumber\\
&+&\left[\frac{D\gamma}{1-\tau(1-3\left\langle x^{2}\right\rangle)}\right]\frac{\partial^{2}}{\partial v^{2}}W(x,v,t)\nonumber\\
&=&-v\frac{\partial}{\partial x}W(x,v,t)-\frac{\partial}{\partial v}\left\{\left[x-x^{3}-\gamma v\right]W(x,v,t)\right\}\nonumber\\
&+&\left[\frac{D\gamma}{1-\tau(1-3\left\langle x^{2}\right\rangle)}\right]\frac{\partial^{2}}{\partial v^{2}}\left\{W(x,t)\right\}\nonumber\\
&-&A\sin(\omega t)\frac{\partial}{\partial v}W(x,t)\nonumber\\
&=&L_{0}W(x,t)+L_{1}W(x,t),
\end{eqnarray}
For convenience, here we consider $\mu=[1-\tau(1-3\langle x^{2}\rangle_{0})]/D$, the steady state probability density is 
\begin{eqnarray}\label{eq:Steadystatedensity}
&&W_{\textrm{st}}(x,v)=Z\exp\nonumber\\
&&\left(-\left[\ln D^{(2)}(x)-\int\int\frac{D^{(1)}(x',v')}{D^{(2)}(x')}dx'dv'\right]\right)\nonumber\\
&=&Z\exp\left(-\ln\frac{\gamma}{\mu}+\frac{\mu}{\gamma}\int\int(x-x^{3}-\gamma v)dxdv\right)\nonumber\\
&=&\frac{Z\mu}{\gamma}\exp\left(\frac{\mu}{\gamma}\left(\frac{1}{2}x^{2}-\frac{1}{4}x^{4}-\frac{\gamma}{2}v^{2}\right)\right).
\end{eqnarray}
Here, the drift and diffusion terms are $D^{(1)}=[x-x^{3}-\gamma v +A\sin(\omega t)]$ and $D^{(2)}=D\gamma/[1-\tau(1-3\langle x^{2}\rangle_{0})]$, respectively. Since the mean of oscillation under the external perturbation is  
\begin{equation}\label{eq:MeansingleDuffing}
\left\langle x_{1}\right\rangle=\int\int x(t)W_{\textrm{ext}}(x,v)dxdv,    
\end{equation}
inserting $W_{\textrm{ext}}(x,v,t)$ into Eq.~(\ref{eq:MeansingleDuffing}), we obtain 
\begin{eqnarray}\label{eq:CalculationsingleDuffing}
\left\langle x_{1}\right\rangle=\int\int xW_{\textrm{ext}}(x,v)dxdv=\int^{t}_{-\infty}R(t-t')F_{\textrm{ext}}(t')dt'.
\end{eqnarray}
As a result, the response function can be recast as 
\begin{eqnarray}\label{eq:resonsefunctionofsingleDuffing}   
R(t-t')&=&\mu\gamma\int_{x}\int_{v}xve^{L_{0}(t-t')}W_{\textrm{st}}(x,v)dxdv\nonumber\\
&=&\mu\gamma\frac{\mathrm{d}}{\mathrm{d}t'}\left\langle x(t)x(t')\right\rangle_{0}.
\end{eqnarray}
Noticeably, the Eq.~(\ref{eq:resonsefunctionofsingleDuffing}) suggests that the response function of the underdamped situation has a same form as the overdamped counterpart. Furthermore, the potential function of the underdamped case is still bistable, we thus consider the correlation function has a same modality as overdameped case as  
\begin{equation}
\left\langle x(t)x(t')\right\rangle_{0}=\left\langle x(t)^{2}\right\rangle_{0} e^{-\lambda_{\textrm{OU}}t'},  
\end{equation}
consequently, the response function is 
\begin{equation}
R(t-t')=\mu\gamma\frac{\mathrm{d}}{\mathrm{d}t'}\left\langle x(t)x(t')\right\rangle_{0}=-\lambda_{\textrm{OU}}\gamma\mu\langle x(t)^{2}\rangle_{0} e^{-\lambda_{\textrm{OU}}t'}.
\end{equation}
After performing a Fourier transform, we have 
\begin{equation}
R(\omega)=-\lambda_{\textrm{OU}}\gamma\mu\langle x(t)^{2}\rangle_{0}\frac{1}{i\omega-\lambda_{\textrm{OU}}},    
\end{equation}
and the signal response can be written as 
\begin{equation}\label{eq:OUsingleDuffing}
\eta_{\textrm{single}}(\omega)=\left[\lambda_{\textrm{OU}}\gamma\mu\langle x(t)^{2}\rangle_{0}\right]^{2}\frac{1}{\omega^{2}+\lambda_{\textrm{OU}}^{2}}.   
\end{equation}

\begin{table*}[htbp]
\renewcommand{\arraystretch}{2.5}
    \centering
    \begin{tabular}{p{30mm}p{80mm}p{30mm}}
     \toprule [1pt]
        Models & Signal response & Noise types \\
     \midrule [1pt]
     
     \multirow{5}{=}{Overdamped bistable model}
     &\multirow{1}{=}{$\eta_{\textrm{single}}=\left(\lambda\frac{\langle x(t)^{2} \rangle_{0}}{D}\right)^{2}\frac{1}{\omega^{2}+\lambda^{2}}$}
     &\multirow{2}{=}{Gaussian white noise} \\
     
     &\multirow{1}{=}{$\eta_{\textrm{meanfield}}=\left(\lambda\frac{\langle x(t)^{2}\rangle_{0}}{D}\right)^{2}\frac{1}{\omega^{2}+\left(c\frac{\langle x(t)^{2}\rangle_{0}}{D}-1\right)^{2}\lambda^{2}}$} \\
     
     &\multirow{1}{=}{$\eta_{\textrm{single}}=\left(\lambda_{\textrm{OU}}\frac{\left\langle x^{2}(t)\right\rangle_{0}+\tau D}{D}\right)^{2}\frac{1}{\omega^{2}+\lambda^{2}_{\textrm{OU}}}$}
     &\multirow{2}{=}{ Ornstein-Uhlenbeck noise } \\
     &\multirow{1}{=}{$\eta_{\textrm{meanfield}}=\left(\lambda_{\textrm{OU}}\frac{\alpha}{D}\right)^{2}\frac{1}{\omega^{2}+\left(c\frac{\alpha}{D}-1\right)^{2}\lambda^{2}_{\textrm{OU}}}$} \\
   \midrule [1pt]   
    \multirow{3}{=}{Underdamped Duffing model} 
     &\multirow{1}{=}{$\eta_{\textrm{single}}=\left(\frac{\lambda\gamma\langle x(t)^{2}\rangle_{0}}{D}\right)^{2}\frac{1}{\omega^{2}+\lambda^{2}}$} 
     &\multirow{2}{=}{ Gaussian white noise } \\
      &\multirow{1}{=}{$\eta_{\textrm{meanfield}}=\left(\frac{\lambda\gamma\langle x(t)^{2}\rangle_{0}}{D}\right)^{2}\frac{1}{\omega^{2}+\left(\frac{c\gamma \langle x(t)^{2}\rangle_{0}^{2}}{D}-1\right)^{2}\lambda^{2}}$} \\
      &\multirow{1}{=}{$\eta_{\textrm{single}}=\left(\lambda_{\textrm{OU}}\gamma\mu\langle x(t)^{2}\rangle_{0}\right)^{2}\frac{1}{\omega^{2}+\lambda_{\textrm{OU}}^{2}}$} 
      &\multirow{2}{=}{Ornstein-Uhlenbeck noise}\\ 
     &\multirow{1}{=}{$\eta_{\textrm{meanfield}}=\left(\lambda_{\textrm{OU}}\gamma\mu\langle x(t)^{2}\rangle_{0}\right)^{2}\frac{1}{\omega^{2}+(c\gamma \mu\langle x(t)^{2}\rangle_{0}-1)^{2}\lambda_{\textrm{OU}}^{2}}$}\\ 
     \bottomrule [1pt]
    \end{tabular}
    \caption{The theoretical signal response formulas for different models, coupling and noise types. The symbol $\langle x(t)\rangle_{0}$ denotes the time average without the external driving, and  $\alpha=\left\langle x^{2}(t)\right\rangle_{0}+\tau D$, $\mu=[1-\tau(1-3\langle x^{2}\rangle_{0})]/D$. We can find that these signal response formulas showcase the typical Lorentz function form. }
    \label{mytab}
\end{table*}
Subsequently, for the case of a nonvanshing coupling, the mean-field Duffing equation can be written as 
\begin{eqnarray}\label{eq:MeanfieldDuffing}
\dot{v}&=&(1-c)x-x^{3}-\gamma v+c\left[X^{0}(t)+ X^{1}(t)\right]\nonumber\\
&+&A\sin(\omega t)+y\nonumber\\ 
\dot{x}&=&v; \dot{y}=-\frac{1}{\tau}y+\frac{1}{\tau}\xi(t).
\end{eqnarray}
The corresponding Fokker-Planck equation of Eq.~(\ref{eq:MeanfieldDuffing}) is
\begin{eqnarray}
\frac{\partial W(x,v,t)}{\partial t}&=&-v\frac{\partial}{\partial x}W(x,v,t)\nonumber\\
&-&\frac{\partial}{\partial v}[\beta+cX^{1}+A\sin(\omega t)]W(x,v,t)\nonumber\\
&+&\left[\frac{D\gamma}{1-\tau(1-3\left\langle x^{2}\right\rangle)}\right]\frac{\partial^{2}}{\partial v^{2}}W(x,v,t)\nonumber\\
&=&-v\frac{\partial}{\partial x}W(x,v,t)+\frac{\gamma}{\mu}\frac{\partial^{2}}{\partial v^{2}}\left\{W(x,t)\right\}\nonumber\\
&-&\frac{\partial}{\partial v}\left\{\beta W(x,v,t)\right\}\nonumber\\
&-&\left[cX^{1}+A\sin(\omega t)\right]\frac{\partial}{\partial v}W(x,t)\nonumber\\
&=&L_{0}W(x,t)+L_{1}W(x,t).
\end{eqnarray}
The steady-state probability density is
\begin{eqnarray}\label{eq:steady-stateprobabilitydensity}
&&W_{\textrm{st}}(x,v)=Z\exp\nonumber\\
&&\left(-\left[\ln D^{(2)}(x)-\int\int\frac{D^{(1)}(x',v')}{D^{(2)}(x')}dx'dv'\right]\right)\nonumber\\
&=&Z\exp\left(-\ln\frac{\gamma}{\mu}+\frac{\mu}{\gamma}\int\int\beta dxdv\right)\nonumber\\
&=&Z\frac{\mu}{\gamma}\exp\left(\frac{\mu}{\gamma}\left[\frac{1-c}{2}x^{2}-\frac{1}{4}x^{4}-\frac{\gamma}{2}v^{2}+cxX^{0}\right]\right).
\end{eqnarray}
Additionally, since the mean oscillation of the Duffing oscillators is
\begin{eqnarray}\label{eq:meanDuffing}
X^{1}(t)=\int\int xW_{\textrm{ext}}(x,v)dxdv
=\int^{t}_{-\infty}R(t-t')F_{\textrm{ext}}(t')dt'.
\end{eqnarray}
Succinctly, the Eq.~(\ref{eq:meanDuffing}) can be recast as 
\begin{equation}\label{eq:meanDuffing1}
X^{1}(t)=\int^{t}_{-\infty}R(t-t')\left[cX^{1}(t)+A\sin(\omega t)\right]dt'.    
\end{equation}
Performing a Fourier transform on Eq.~(\ref{eq:meanDuffing1}), then we have 
\begin{equation}
\tilde{X}_{1}(\omega)=\chi(\omega)\left(c\tilde{X}_{1}(\omega)+F(\omega)\right),    
\end{equation}
and the resulting relation
\begin{equation}
\tilde{X}_{1}(\omega)=\frac{\chi(\omega)}{1-c\chi(\omega)}F(\omega)=\tilde{\chi}(\omega)F(\omega).    
\end{equation}
Resembling the vanishing coupling situation, we can obtain the response function as
\begin{eqnarray}
R(t-t')&=&\mu\int_{x}\int_{v}xve^{L_{0}(t-t')}W_{\textrm{st}}(x,v)dxdv\nonumber\\
&=&\mu\gamma\frac{\mathrm{d}}{\mathrm{d}t'}\left\langle x(t)x(t')\right\rangle_{0}\nonumber\\
&=&-\mu\gamma\lambda_{\textrm{OU}}\gamma \langle x(t)^{2}\rangle_{0} e^{-\lambda_{\textrm{OU}}t'}.
\end{eqnarray}
Furthermore, after the Fourier transform, the response function is recast as
\begin{equation}
\chi(\omega)=-\mu\lambda_{\textrm{OU}}\gamma\langle x(t)^{2}\rangle_{0} \frac{1}{i\omega-\lambda_{\textrm{OU}}},    
\end{equation}
which leads to the form
\begin{equation}
\tilde{\chi}(\omega)=\frac{\chi(\omega)}{1-c\chi(\omega)}=\frac{\mu \lambda_{\textrm{OU}}\gamma\langle x(t)^{2}\rangle_{0} }{\left\{1-c\gamma\mu\langle x(t)^{2}\rangle_{0}\right\}\lambda_{\textrm{OU}}-i\omega}.
\end{equation}
Finally, the signal response can be written as 
\begin{equation}\label{eq:MeanfieldDuffingOU}
\eta_{\textrm{meanfield}}(\omega)=\left[\lambda_{\textrm{OU}}\gamma\mu\langle x(t)^{2}\rangle_{0}\right]^{2}\frac{1}{[1-c\gamma\mu \langle x(t)^{2}\rangle_{0} ]^{2}\lambda_{\textrm{OU}}^{2}+\omega^{2}}.
\end{equation}
As Fig. ~\ref{Fig:3} (a) shows, we can still find that the mean-filed coupling can enhance the signal response for both the Gaussian white and the OU noise situations. Nevertheless, unlike the reduction of the signal response induced by the noise color in the overdamped systems, the theoretical signal response for the OU noise case is larger than that of the Gaussian white counterpart in the underdamped systems. Furthermore, such a phenomenon of the colored noise enhanced signal response is significant when the driving is sufficiently low, see Fig. ~\ref{Fig:3} (b).

On the one hand, a theoretical understanding of the coupled Duffing oscillators driven by colored noise is still insufficient. On the other hand, the issue concerning whether colored noise outperforms white noise in weak signal response needs more theoretical evidence from different viewpoints and situations. The main results in this section demonstrate theoretically that colored noise can be better than white noise in weak signal response in Duffing systems.

Additionally, as Table.~\ref{mytab} shows, the theoretical signal response expressions of systems ranging from a single overdamped bistable oscillator driven by Gaussian white noise to the mean-field coupled Duffing oscillators perturbed by OU noise fully showcase the explicit Loren\textcolor[rgb]{1.00,0.00,0.00}{t}z function form.

\section{Influence of diverse factors on SR}\label{Sec2C}
For a brief summary, based on LRT, we found that the external driving frequency, coupling, and noise color all have significant influence on SR.  Firstly, as the external driving frequency increases, the signal response curves in diverse cases fully decrease within the form of the inverse-Sigmoid-curve, which demonstrates that a small enough driving frequency is a prerequisite for the emergence of SR.

Secondly, compared to the single particle situation, the signal response of the mean-field coupled counterparts are significantly improved, especially for the cases of the low-frequency signals.

Thirdly, for the overdamped bistable model, we found that although the optimal signal response of the OU noise is lower than that of the Gaussian white noise, the signal response of the OU noise case for a given degree of noise can be better than that of the white one if the driving frequency is sufficiently low. However, for the underdamped Duffing model, we demonstrated that the OU noise outperforms the Gaussian white one from both viewpoints of the optimal signal response and the signal response for a fixed noise intensity.

As one application of the general expression, we further proposed the inherent mechanisms for the influence of diverse factors, like the external signal frequency, mean-field coupling, and the noise color (namely the self-correlation time of noise), on signal response based on the general expression in Table.~\ref{mytab}.
\subsection{The external driving frequency}
As shown in Table.~\ref{mytab}, if we concentrate on the influence of the external driving frequency, the signal response can be considered as a typical Lorentz function, as a result, the signal response profiles in diverse situations fully showcase the inverse-Sigmoid-trend as the driving frequency increases.  
\subsection{The mean-field coupling}
Taking the mean-field coupling into consideration, we can concisely
compare the signal responses of the single overdamped particle and its mean-filed coupled counterpart. As the formulas shown in the first two rows of Table.~\ref{mytab}, we can find that the effect of mean-field coupling on signal response is concentrated in the quadratic function $f(c)=(c\langle x(t)^{2}\rangle_{0}/{D}-1)^{2}$, in which the term $\langle x(t)^{2}\rangle_{0}/{D}$ is a constant. When the coupling is vanished, $f(0)=1$, the signal response is equal to one of the single particle case. Furthermore, when the coupling gradually increases, $f(c)$ decreases firstly and then increases, and consequently, the mean-field coupling can enhance SR in a bell-shaped form. Similar conclusions can be obtained in the other three-pairwise situations in Table.~\ref{mytab}. 

\subsection{The noise color}
To reveal the role of the noise color on SR, we here first compare the signal responses of the mean-field coupled overdamped bistable oscillators driven by Gaussian white noise to that of the OU noise, as shown in the second and the fourth rows in Table.~\ref{mytab}. For convenience, we consider the optimal signal response as $\eta^{\textrm{optimal}}$ as the magnitude of SR. When the noise-evoked hopping rate matches the external driving force, i.e., $\lambda_{\textrm{OU}}=2\omega/\pi$, we can obtain the optimal noise intensity as 
\begin{equation}
D_{\textrm{optimal}}=\frac{1}{4}\left[ \ln\left( \frac{1-\frac{3}{2}\tau}{\sqrt{2}\omega}\right)\right]>D_{\textrm{optimal}}(\tau=0).    
\end{equation}
Therefore, the optimal noise intensity is increased as the noise color increases. Additionally, inserting the optimal noise intensity $D_{\textrm{optimal}}$ and the escape rate $\lambda_{\textrm{OU}}$ into the theoretical signal response of the mean-field coupled overdamped oscillators driven by OU noise, we can get 
\begin{equation}
\eta^{\textrm{optimal}}_{\textrm{over}}\approx \frac{(\tau D_{\textrm{optimal}}+1)^{2}}{(2c\tau +3)D_{\textrm{optimal}}^{2}+2cD_{\textrm{optimal}}}.
\end{equation}
Since the coupling $c=0.1$ are small, and the variable $\tau D_{\textrm{optimal}}\ll 1$ for small correlation time, 
the optimal signal response can be further recast as  
\begin{equation}
\eta^{\textrm{optimal}}_{\textrm{over}}\approx\frac{1}{(2c\tau+3)D_{\textrm{optimal}}^{2}}. 
\end{equation}
As a result, the magnitude of SR is reduced as the noise color increases. The mechanism behind the phenomenon of noise-color-reduced SR is that the noise color weakens the escape rate. 

On the other hand, for the situation of the mean-field coupled Duffing elements disturbed by OU noise, we can obtain 
\begin{equation}\label{eq:Duffingdenominator}
\eta^{\textrm{optimal}}_{\textrm{under}}\approx\frac{1}{\frac{1+\frac{\pi^{2}}{4}}{\gamma^{2}(1+2\tau)^{2}}D_{\textrm{optimal}}^{2}-\frac{2c}{\gamma(1+2\tau)}D_{\textrm{optimal}}+c^{2}}.   
\end{equation}
For convenience, here we consider the denominator of the Eq.~(\ref{eq:Duffingdenominator}) as function $f(D_{\textrm{optimal}})$. The axis of symmetry of a quadratic equation with one variable can be written as
\begin{equation}
D_{\textrm{optimal}}=\frac{2c\gamma}{2+0.5\pi^{2}}(1+2\tau)\propto 1+2\tau.
\end{equation}
Compared to the white noise situation, when $\tau>0$, the axis of symmetry of $f(D_{\textrm{optimal}})$ moves rightward, as a consequence, a new region that corresponds to the decreasing trend of $f(D_{\textrm{optimal}})$ emerges.
Thereby, for a nonvanishing correlation time, the optimal signal response curve is bell-shaped.
In summary, the phenomenon of the nonmonotonical enhancement of SR in Duffing oscillators is due to the complex interaction of the damping, coupling and the noise color.

\section{Conclusions}\label{Sec2D}

In conclusion, we have theoretically investigated the signal response of both overdamped and underdamped bistable systems with respect to two types of typical noise and the weak sinusoidal signal based on linear response theory. These theoretical results capture the major conclusions in SR in bistable systems, like, the mean-filed coupling can significantly enhance the SR, the signal response monotonically reduces as the inverse-
Sigmoid-curve as the external driving frequency rises, the noise color weakens the SR in the overdamped bistable system but strengthens the underdamped Duffing counterpart and so on. Furthermore, we found that the theoretical signal responses in diverse situations can be concluded into a uniform expression that possesses the Loren\textcolor[rgb]{1.00,0.00,0.00}{t}z function form. As one direct application of the general signal response, we clearly proposed the role of distinct factors, for instance, the external driving frequency, the mean-field coupling, and the noise color, in this resonance-like signal response.  These results integrate the independent theoretical results and thus contribute to a deeper understanding of stochastic resonance in bistable systems.

Since bistable systems have served as a testbed to investigate the resonant signal response in fields ranging from physics and neuroscience to engineering and climate. The main conclusions originating from the eight typical situations in the present work are far from covering all the conventional scenarios, and to the best of our knowledge, two canonical instances not included are the quantum stochastic resonance~\cite{Lofstedt1994,Wagner2019} and the entropic stochastic resonance~\cite{Burada2008,HPZhang2022}. Notwithstanding, our results provide a seemingly plausible guess that the stochastic resonance-like signal response in bistable systems may be capable of an uniform theoretical expression.

\section{Acknowledgments}
Cong Liu warmly thanks professors Xiaoming Liang and Xiyun Zhang for the insightful suggestions.  
This work is funded by Science Research Project of Hebei Education Department under Grant No. QN2025065 and Science Foundation of Hebei Normal Universty under Grant No. L2025B10.
We acknowledge financial support from the National Natural Science Foundation of China (Grants No. 12375032 and No. 12247101), and from the Fundamental Research Funds for the Central Universities and "111 Center" (Grant No. lzujbky-2024-jdzx06 and No.B20063). This work was partly supported by the Natural Science Foundation of Gansu Province (No. 22JR5RA389) and Longyuan-Youth-Talent Project of Gansu Province.
We thank the anonymous reviewers for critical comments that helped improve the paper.
\section*{AUTHOR DECLARATIONS}
\subsection*{Conflict of Interest}
The authors declared that they have no conflicts of interest to this work.
\section*{Data Availability Statement}

The data that support the findings of this study is available within the article
\appendix*
\section{}
In this Appendix, we present the detailed calculations concerning the average motion and the response function for the Duffing model. For the case of the single Duffing driven by the Gaussian white noise as an example, the mean of oscillation under the external perturbation is 
\begin{equation}\label{eq:MeansingleDuffing123}
\left\langle x_{1}\right\rangle=\int\int x(t)W_{\textrm{ext}}(x,v)dxdv,    
\end{equation}
inserting $W_{\textrm{ext}}(x,v,t)$ into Eq.~(\ref{eq:MeansingleDuffing123}), we obtain 
\begin{eqnarray}\label{eq:CalculationsingleDuffing}
&&\left\langle x_{1}\right\rangle=\int\int xW_{\textrm{ext}}(x,v)dxdv\nonumber\\
&=&\int_{x}\int_{v}\int_{-\infty}^{t}xe^{L_{0}(t-t')}L_{1}(t')W_{\textrm{st}}(x,v)dt'dxdv\nonumber\\
&=&\int_{-\infty}^{t}\left[\int_{x}\int_{v}xe^{L_{0}(t-t')}L_{1}(t')W_{\textrm{st}}(x,v)dxdv\right]dt'\nonumber\\
&=&\int_{-\infty}^{t}F_{\textrm{ext}}(t')\left[\int_{x}\int_{v}xe^{L_{0}(t-t')}\frac{1}{D}\gamma vW_{\textrm{st}}(x,v)dxdv\right]dt'\nonumber\\
&=&\frac{\gamma}{D}\int_{-\infty}^{t}F_{\textrm{ext}}(t')\left[\int_{x}\int_{v}xve^{L_{0}(t-t')}W_{\textrm{st}}(x,v)dxdv\right]dt'\nonumber\\
&=&\int^{t}_{-\infty}R(t-t')F_{\textrm{ext}}(t')dt'.
\end{eqnarray}
As a result, the response function can be expressed as 
\begin{eqnarray}\label{eq:resonsefunctionofsingleDuffing1}   
R(t-t')&=&\frac{\gamma}{D}\int_{x}\int_{v}xve^{L_{0}(t-t')}W_{\textrm{st}}(x,v)dxdv\nonumber\\
&=&\frac{\gamma}{D}\int_{x}\int_{v}\int_{x'}\int_{v'}xv'P(x,v,t|x',v',t')\nonumber\\
&\times& P_{0}(x',v')dxdvdx'dv'\nonumber\\
&=&\frac{\gamma}{D}\int_{x}\int_{v}\int_{x'}\int_{v'}x\frac{\mathrm{d}}{\mathrm{d}t'}\left[x'P(x,v,t|x',v',t')\right]\nonumber\\
&\times&P_{0}(x',v')dxdvdx'dv'\nonumber\\
&-&\frac{\gamma}{D}\int_{x}\int_{v}\int_{x'}\int_{v'}xx'\frac{\mathrm{d}}{\mathrm{d}t'}\left[P(x,v,t|x',v',t')\right]\nonumber\\
&\times& P_{0}(x',v')dxdvdx'dv'\nonumber\\
&=&\frac{\gamma}{D}\int_{x}\int_{v}\int_{x'}\int_{v'}\frac{\mathrm{d}}{\mathrm{d}t'}\left[xx'P(x,v,t|x',v',t')\right]\nonumber\\
&\times& P_{0}(x',v')dxdvdx'dv'\nonumber\\
&+&\frac{\gamma}{D}\int_{x}\int_{v}\int_{x'}\int_{v'}xx'e^{L_{0}(t-t')}L_{0}(t-t')\nonumber\\
&\times& P_{0}(x,v)dxdvdx'dv'\nonumber\\
&=&\frac{\gamma}{D}\frac{\mathrm{d}}{\mathrm{d}t'}\int_{x}\int_{v}\int_{x'}\int_{v'}xx'P(x,v,t|x',v',t')\nonumber\\
&\times&dxdvdx'dv'\nonumber\\
&=&\frac{\gamma}{D}\frac{\mathrm{d}}{\mathrm{d}t'}\left\langle x(t)x(t')\right\rangle_{0}.
\end{eqnarray}
Furthermore, the procedures for deducing the response function and mean oscillation for the case of a single Duffing particle driven by the OU noise are similar to the white noise situation shown above.
\section*{Reference}
\nocite{*}
\bibliography{aipsamp}

\end{document}